\documentstyle[12pt,aaspp4,psfig]{article}
\newcommand{\etal}{{\it et al. \,}}
\newcommand{\grados}{^\circ}
\lefthead{Aparicio \etal}
\righthead{DDO 187}

\begin{document}

\title{DDO 187: do dwarf galaxies have extended, old
halos?\footnote{Based on observations made with the 2.5 m Nordic Optical
Telescope operated on the island of La Palma by NOT S.A. in the Spanish
Observatorio del Roque de Los Muchachos of the Instituto de Astrof\'\i
sica de Canarias.}}

\author{Antonio Aparicio} 
\affil{Instituto de Astrof\'\i sica de Canarias,
E38200 - La Laguna, Tenerife, Canary Islands, Spain}

\author{Nikolay Tikhonov and Igor Karachentsev}
\affil{Special Astrophysical Observatory, Stavropol, Russia}

\begin{abstract}

If dwarf galaxies are primeval objects in the Universe, as hierarchical
galaxy formation scenarios predict, they should show traces of their old
stellar populations, perhaps distributed in extended, differentiated
structures. The working hypothesis that such a structure could exist is
tested for the case of DDO 187, a field dIrr galaxy showing a high gas
fraction and  low metallicity. For this purpose, the structure, star
formation history (SFH) and other properties of the galaxy are analyzed
using the spatial distribution of stars, the color--magnitude diagrams (CMDs)
of about 1500 resolved stars and the fluxes of H~{\sc ii} regions, together
with data about the gas distribution.

From the $I$ magnitude of the tip of the red giant branch ($I _{\rm
TRGB}$), the distance of DDO 187 to the Milky Way is estimated to be
$2.5\pm 0.2$ Mpc. The distance to several neighbor galaxies and
groups has been computed, showing that DDO 187 is probably an
isolated, field galaxy.

The distance of DDO 187 to the Milky Way is almost three
times smaller than that obtained from Cepheid
light-curves. Considering that this is the third case in which such a
large disagreement is detected, it seems clear that Cepheid distance
estimates based on a few stars, as usually happens in dwarf galaxies,
must be accepted with caution.

The star formation history of DDO 187 has been analyzed. The central
region of DDO 187 shows an overall time decreasing star formation rate
with a strong burst in its central region which happened between 20
and 100 Myrs ago and a present-day star formation activity three times
smaller than the maximum one. Besides this, a spatially extended
stellar component has been found that has no young stars and exceeds
the size of the gas component.

In short, several results suggest that DDO 187 has a two-component
halo/disk-like structure: (i) differentiated morphologies for the inner
(flat) and outer (spheroidal) stellar components; (ii) a gas component
less extended than the outer stellar component, and (iii) an outer component
lacking young stars, which are abundant in the inner component. The
working hypothesis that a real halo/disk structure could be present is
discussed. The conclusion is reached that the two-component hypothesis
is not unrealistic, but nothing can be definitely stated until more
detailed data, ideally including kinematics, are available.

\end{abstract}


\section{Introduction}

In hierarchical scenarios of galaxy formation (White \& Rees 1978),
dwarfs would be the first galaxies to be formed, from small-amplitude 
density fluctuations and would latter on merge  to
form larger galaxies and structures. The fact that the gas-poor, dEs
are usually found in denser parts of the Universe than the gas-rich
dIrrs may indicate that both have the same origin, the former being in
a more evolved evolutionary stage and in the process of being soon
incorporated into bigger neighboring galaxies. If this were the case,
isolated dIrrs like DDO 187 should still have the traces of their
formation process and early evolution and, eventually, a similar
underlying structure to that of the less distorted dEs. It is hence of major
relevance to study the structure, kinematics and spatial distribution
of stellar populations, including the oldest ones, of these small, gas
rich galaxies, checking in particular whether they have old/young
subsystems similar to the halo/disk ones of spirals or any other
structures that would trace their formation scenario. The
main purpose of this paper is to test the hypothesis of the existence of a
two-component, halo/disk structure in the field dIrr galaxy DDO 187.

DDO 187 (UGC 9128) is a resolved, gas-rich dwarf galaxy first cataloged by van
den Bergh (1959). It shows a morphology typical of dIrr galaxies, with
the bluest stellar population lying on one-half of its body, and has
been subject of several works directed toward the study of its
distance, stellar content, neutral gas, and chemical
abundances. Aparicio, Moles, \& Garc\'\i a-Pelayo (1988) showed a
color--magnitude diagram (CMD) and roughly estimated the distance to
the galaxy to be $4.4\pm 1.0$ Mpc, based on the magnitude of the
brightest stars. Skillman, Kennicutt, \& Hodge (1989) found a very low
oxygen abundance of $12+\log {\rm O/H}=7.36$. Recently, van Zee , Haynes, \& Salzer (1997a) have obtained $12+\log {\rm O/H}=7.75$ and 7.73 for
two H~{\sc ii} regions in the galaxy.  Lo, Sargent, \& Young (1993)
determined its heliocentric H~{\sc i} velocity (153 km s$^{-1}$) and
using a distance of 4 Mpc, obtained its H~{\sc i} and total virial
masses to be $M_{\rm H}=5\times10^7$ M$_\odot$ and $M_{\rm
VT}=9.8\times10^7$ M$_\odot$. Van Zee, {\it et al.} (1997b),
assuming a distance of 3.4 Mpc, obtained $M_{\rm H}=3.5\times10^7$
M$_\odot$. Using surface photometry, Patterson \& Thuan (1996) derived
an exponential scale length of 0.25 kpc, for a distance of 4.4 Mpc.
Recently Hoessel, Saha, \& Danielson (1998) presented the results of
their time-extended research of variable stars in DDO 187. They find
12 variable stars, and identify two of them as very likely Cepheids
and three more as possible Cepheids. Using their periods and
luminosities, they derive a distance of $7.0\pm1.2$ Mpc for DDO 187.

We present new ground-based $V$, $I$, and $H_\alpha$ photometry of
DDO~187 and analyze the properties of this galaxy. The paper is
organized as follows. Observations are described in Section 2. The CMD of
the resolved stars is presented in Section 3. Our photometry reach the
tip of the red giant branch (TRGB), allowing an independent
determination of the distance, which does not agree with the
Cepheid-based one. This is discussed in Section 4. The ionized
distribution of H and the properties that can be inferred for the
ionizing stars are discussed in Section 5. Section 6 is devoted to the
study of the structure,  star formation history (SFH), and 
spatial distribution of stellar populations, discussing, in
particular, the hypothesis that DDO 187 had a two-components
(halo/disk-like) structure. In Sec. 7, several global properties of
the galaxy are given. Finally, Sec. 8 summarizes the main results of
the paper.

\section{Observations and data reduction}

\placetable{journal}

\begin{figure}
\centerline{\psfig{figure=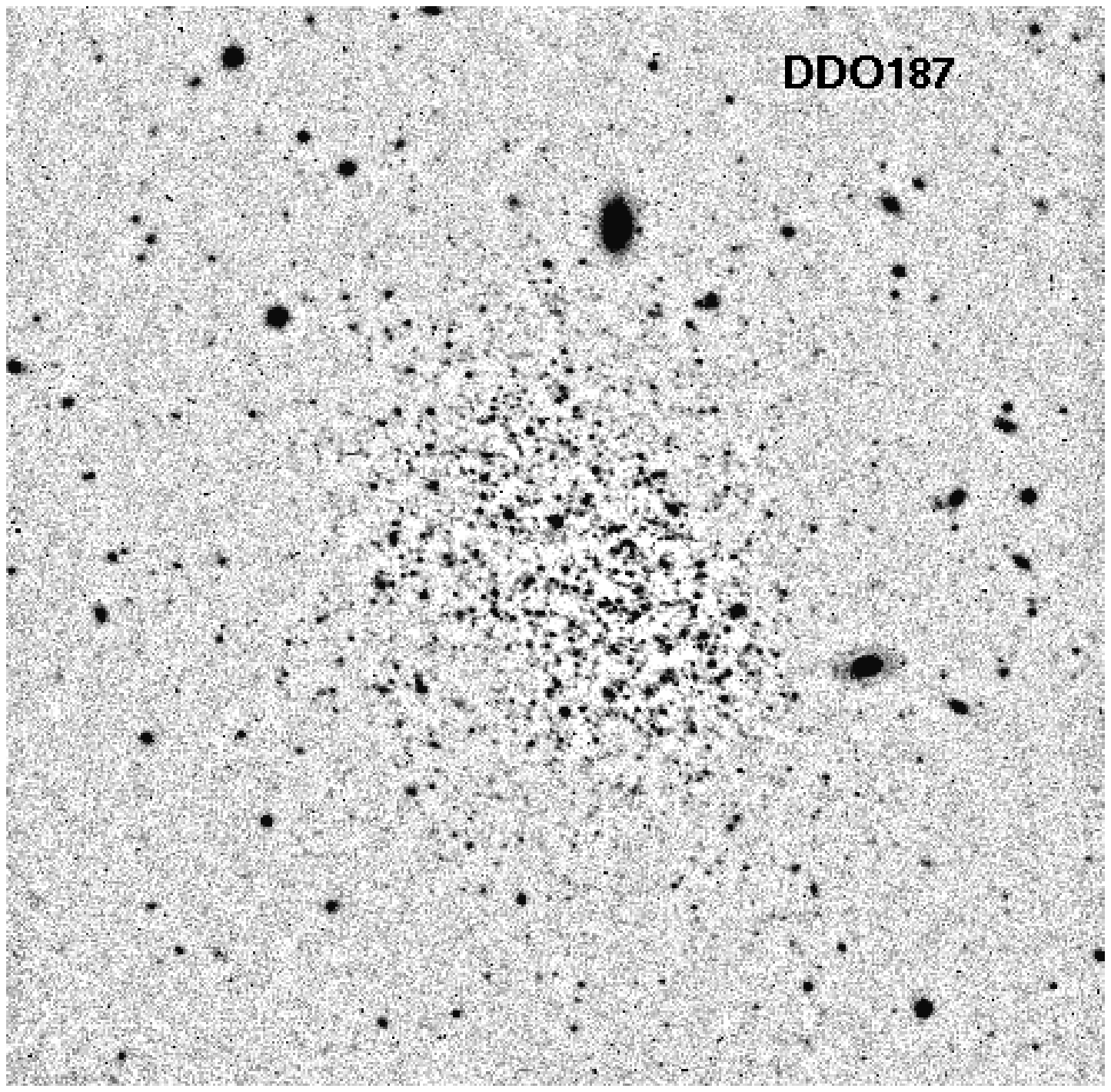,width=16cm}}
\figcaption[ima.ps]{$I$ image of DDO 187. The total field is
$3.75\times3.75(')^2$ and the integration time 900 s. North is up, East
is left.
\label{ima}}
\end{figure}

Images of DDO 187 were obtained in the $V$ and $I$ Johnson--Cousins
filters and in a narrow-band $H_\alpha$ filter with the NOT (2.5 m) at
Roque de los Muchachos Observatory on the island of La Palma  (Canary Islands,
Spain). The HiRAC camera was used with a $2048\times 2048$ Loral CCD
binned to $2\times2$. After binning it provides a scale of 0.22 $''$/pix
and a total field of $3.75\times 3.75$  arcmin$^2$. Total integration times
were 3600 s in $V$, 3000 s in $I$, and 1200 s in $H_\alpha$. A 400 s
exposure was also taken with an $H_\alpha$-continuum filter. Moreover,
integrations of 1200 s in $V$ and 1000 s in $I$ were made of a nearby
field to correct the foreground contamination of the broad-band
images. Observations of DDO 187 were taken with seeing always in the range
$0\farcs7-0\farcs9$. Table \ref{journal} gives the journal of
observations.  Figure \ref{ima} shows one of the $I$ band images of DDO
187. A beautiful picture of the galaxy can be seen in Hoessel {\it et
al.} (1998).

\placetable{journal}
\placefigure{ima}

Bias and flat-field corrections were done with IRAF. Then, DAOPHOT and
ALLSTAR (Stetson 1994) were used to obtain the instrumental photometry
of the stars. Eighteen standard stars from the list of Landolt (1992)
were measured during the observing run to calculate the atmospheric
extinctions for each night and the equations transforming into the Johnson--Cousins standard photometric
system. A total of about 180 measures in both the  $V$ and $I$ bands of these
standards were used. The transformation equations are:

\begin{equation}
(V-v)=25.205-0.106(V-I); ~~~~\sigma=0.005
\end{equation}
\begin{equation}
(I-i)=24.498+0.011(V-I); ~~~~\sigma=0.004
\end{equation}

\noindent where capital letters stand for Johnson--Cousins magnitudes and
lower-case letters refer to instrumental magnitudes corrected for
atmospheric extinction. The $\sigma$ values are the dispersions of the
fits at the centers of mass of the point distributions;
hence they are the minimum internal zero-point errors. Dispersions of the
extinctions for each night vary from $\sigma=0.009$ to $\sigma=0.019$
and dispersions of the aperture corrections are of the order of
0.04. Putting all these values together, the total zero-point error of
our photometry can be estimated to be about 0.05 for both bands.

One standard star from Oke (1990) was observed three times to calibrate the
$H_\alpha$ images. The dispersion of the measures was better than 0.01 
magnitudes.

DAOPHOT and ALLSTAR provide with PSF model to star profile fitting
errors. In general, they do not reproduce the external errors of the
photometry, coming mainly from stellar crowding and blending (see
Aparicio, \& Gallart 1995 and Gallart, Aparicio, \& V\'\i lchez
1996a), but they provide with an indication of the internal accuracy
of the photometry. In this sense, our average ALLSTAR errors are 0.01,
0.04 and 0.16 for $I=20$, 22 and 24 respectively and for $V=21.2$,
23.3 and 25.4, respectively.

\section{The color--magnitude diagram}

\begin{figure}
\centerline{\psfig{figure=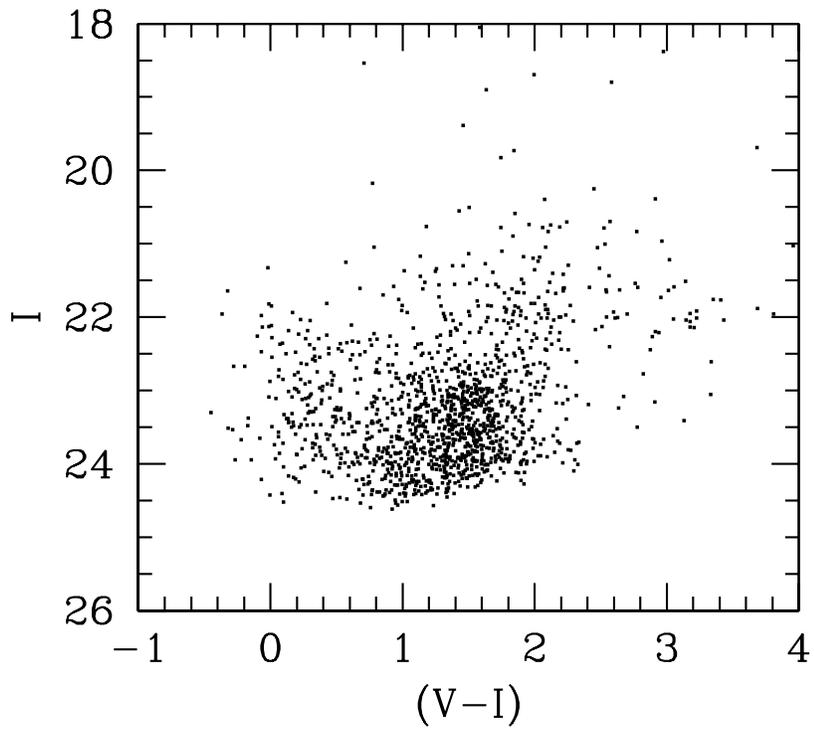,width=16cm}}
\figcaption[cmd.eps]{CMD of DDO 187.
\label{cmd}}
\end{figure}

\placefigure{cmd}

The CMD of DDO 187 is shown in Figure \ref{cmd}. It is typical of a
resolved galaxy with recent star formation (compare, for example, with the
richer CMD of NGC 6822 by Gallart et al. 1996a). A
well populated structure is visible in the region $I\geq 23$, $1.0\leq
(V-I)\leq 1.8$. It corresponds to the {\it red-tangle} (see Aparicio
\& Gallart 1994; Aparicio {\it et al.} 1996). It is formed by low-mass
stars in the RGB and AGB phases and it is hence the trace of an
intermediate-age to old stellar population. A well populated
blue-plume is also visible at $I\geq 21.5$, $(V-I)\leq 0.8$. Formed by
intermediate to massive stars in the MS and/or the core He-burning
blue-loop phases, it is the trace of significant recent star
formation, in good agreement with the presence of H~{\sc ii} regions in the
galaxy. At least some of the red stars in the region $20.5\leq I\leq
23$, $(V-I)\geq 1.6$ are likely AGB stars forming a {\it red-tail}
structure of the kind discussed by Aparicio \& Gallart (1994).

\begin{figure}
\centerline{\psfig{figure=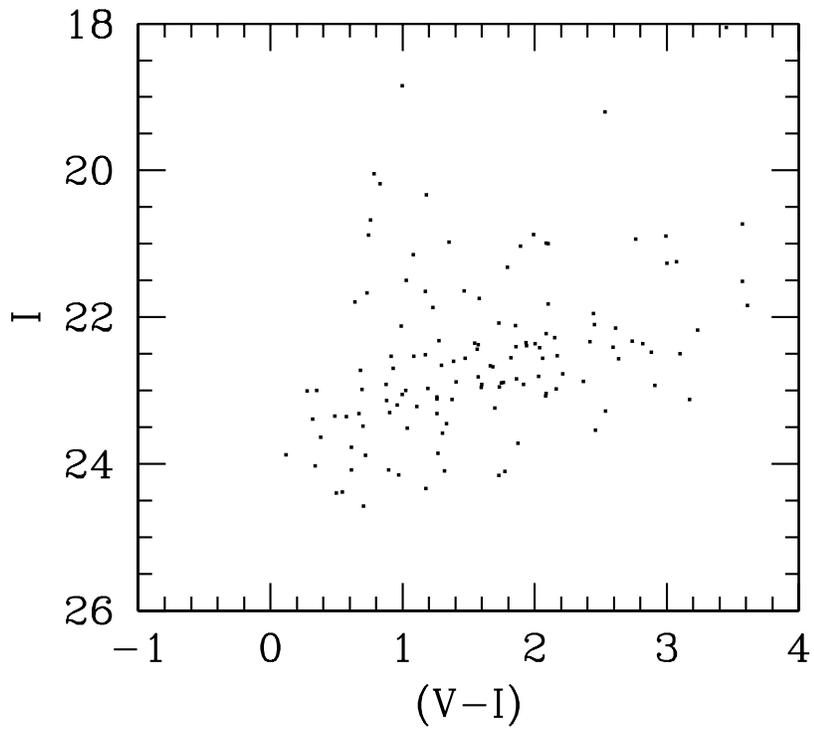,width=16cm}}
\figcaption[f_cmd.eps]{CMD of a nearby companion field mapping 
foreground contamination.
\label{f_cmd}}
\end{figure}

\placefigure{f_cmd}

Figure \ref{f_cmd} shows the CMD of a companion field, close to DDO
187 but far enough to guarantee that it contains no stars of the
galaxy. The distribution of stars in this diagram indicates that most
stars brighter than $I\sim 21$ and some of the fainter
 in the CMD of Fig. \ref{cmd} are foreground stars
belonging to the Milky Way. The CMD of Fig. \ref{f_cmd} will be used
to correct for foreground contamination the star counts in DDO 187
used in Section 6.

In summary, the CMD of DDO 187 indicates that it has
had an extended SFH, as evidenced by the presence of young blue stars
and a red-tangle of intermediate-age and/or old stars. A quantitative
estimate of the SFH is given in Section 6.2.

\section{The distance of DDO 187}

\subsection{The distance from the TRGB}

Hoessel {\it et al.} (1998) made a wide time coverage survey of variable
stars in DDO 187. They identified two stars in the galaxy as very likely
Cepheids and three more as possible Cepheids. Using them, they determine a
distance of $7.0\pm1.2$ Mpc. This distance implies that the TRGB of DDO 187
should be fainter than $I\sim 25$, and hence difficult to reach by usual
ground-based telescopes.

However, this seems not to be the case in the light of the CMD shown in
Figure \ref{cmd}. An independent estimate of the distance to DDO 187 can
be made using the magnitude of the TRGB and the relationships of Lee,
Freedman, \& Madore (1993). We have determined the magnitude of the TRGB 
from the CMD of Fig. \ref{cmd}, applying a Sobel filter to the luminosity
function (LF) of stars with colors in the interval $1.3\leq (V-I)\leq
1.7$, and obtained $I_{\rm TRGB}=22.95\pm 0.17$. The error has been
estimated from the width of the peak produced by the Sobel filter around
the TRGB. Several binning sizes and initial values for the LF were
used, the former result being the average of all them. 

Crowding, blending and other observational effects modify the measured
star magnitudes in the sense that, in average, they appear brighter
than they are (see a discussion of this in Aparicio, \& Gallart 1995
and Gallart et al. 1996a). This would eventually affect the magnitude
of the TRGB. To check this effect and for the analysis of the stellar
populations presented in Section 6, we have performed crowding tests
using artificial stars in the way described by Stetson (1994). We have
added 1190 artificial stars to our images with initial magnitude and
color $I=23.0$, $(V-I)=1.5$, roughly corresponding to the position of
the TRGB. 90\% of them were recovered, their average magnitude and
color being $I=23.01\pm 0.19$ and $(V-I)=1.48\pm 0.14$, showing that
shifts produced by observational effects are negligible at the
position of the TRGB.

The extinction towards DDO 187 can be estimated to be $E(B-V)=0.005$
(Burstein \& Heiles 1982). This can be neglected and hence the former
value for $I_{\rm TRGB}=22.95\pm 0.17$ can be assumed to be free of
extinction and observational effects.

To apply the relationships of Lee {\it et al.} (1993), the metallicity
of the stars and the bolometric correction (BC) at the TRGB are
necessary. Both can be estimated from the color indices of RGB
stars. $(V-I)_{\rm TRGB}=1.53\pm 0.03$ and $(V-I)_{-3.5}=1.45\pm 0.04$
have been obtained as the median of the color indices of stars with
$22.95\leq I\leq 23.05$ and $23.4\leq I\leq 23.5$ respectively and
having colors in the interval $1.0\leq (V-I)\leq 2.0$. Some 35 stars
have been used in each case and the quoted errors have been simply
obtained as $\sigma(n-1)^{-1/2}$ for each sample. From $(V-I)_{-3.5}$,
an average metallicity of [Fe/H] $=-1.36\pm 0.11$ is found using the
calibration by Lee {\it et al.} (1993). This Fe abundance corresponds
to $Z=0.0008\pm 0.0002$, which is similar to the values obtained from
the oxygen abundance of the H~{\sc ii} regions by Skillman {\it et
al.} (1989) and by van Zee {\it et al.} (1997a) ($Z=0.0006$ and
$Z=0.001$, respectively).

These values can finally be introduced in the relationships of Lee {\it et
al.} (1993) to obtain the absolute magnitude of the TRGB, $M_{I,\rm
TRGB}=-4.06$, and hence a distance modulus of $(m-M)_0=27.0\pm 0.2$,
which corresponds to a distance to the Milky Way of $d_{\rm MW}=2.5\pm
0.2$ Mpc.

\subsection{The distance of DDO 187 to the closest galaxies and
groups}

To check whether DDO 187 can be considered a field galaxy, its distance
to other nearby galaxies has been computed. The selected galaxies are
GR 8, DDO 187, Andromeda, M81, M94 and M101, as well as the barycenter
of the Local Group. The first two galaxies are apparently the closest
dwarfs to DDO 187. The other ones are the dominant members of the
closest groups. The distances to the Milky Way adopted for each one
are: GR 8, 2.2 Mpc (Dohm-Palmer {\it et al.}  1998); DDO 190,
2.9 Mpc (Aparicio {\it et al.} 1999); Andromeda, 0.77 Mpc (Freedman \&
Madore 1990); M 81, 3.6 Mpc (Freedman {\it et al.}  1994); M 94, 4.7
Mpc (Mulder \& van Driel 1993, adopting $H_0=70$ kms$^{-1}$Mpc$^{-1}$)
and M 101, 7.0 Mpc (Stetson {\it et al.} 1998).  The position of the
barycenter of the Local Group has been calculated neglecting the masses
of all the galaxies in the LG except Andromeda and the Milky Way,
assuming that the the mass of the Milky Way is 0.7 times that of
Andromeda (Peebles 1989) and adopting a value of 0.77 Mpc for the
Milky Way--Andromeda distance (Freedman \& Madore 1990).

In this way, the resulting distances of DDO 187 to each galaxy are: 
0.85 Mpc to GR 8; 1.06 Mpc to DDO 190; 2.9 Mpc to Andromeda; 3.2
Mpc to M 81; 2.6 Mpc to M 94; 5.2 Mpc to M101 and 2.7 Mpc to the
barycenter of the Local Group. Hence DDO 187 is likely a field galaxy
lying between the Local, M81 and M94 groups, the closest galaxy being
the dIrr GR 8. All the distances are summarized in Table \ref{neig}.

\placetable{neig}

\subsection{The Cepheid distance problem: comparison of Cepheid and
TRGB distances}

\placetable{dist}

The large disagreement between the Cepheid and TRGB distances merits
some comment. While the Cepheid P--L relationship is the best distance
indicator for nearby galaxies, its reliability has been recently
questioned by several authors (Aparicio 1994; Saha {\it et al.} 1996;
Tolstoy {{\it et al.}} 1998). The problem arises from the fact that,
for several dwarf galaxies (namely, Pegasus, Leo A and, now, DDO 187),
a large disagreement has been detected between Cepheid based distances
and TRGB or red-clump (RC) based distances. Table \ref{dist} summarizes the
distances found with each method for the three galaxies. Column 1
gives the galaxy name; columns 2 to 4 list the Cepheid distance, the
number of Cepheids used to determine it and the reference; columns 5 to
7 give the CMD-based distance, the feature (TRGB or RC) used to
estimate it and the reference. In all cases, Cepheid distances were
based on just a few stars showing light curves resembling those of
bona-fide Cepheids. However, latter accurate two-color photometry
(Aparicio 1994; Tolstoy {{\it et al.}} 1998) demonstrated that the
colors of these stars could in no way be reconciled with those of
Cepheid stars, which is also stressed by Saha {\it et al.} (1996).

\begin{figure}
\centerline{\psfig{figure=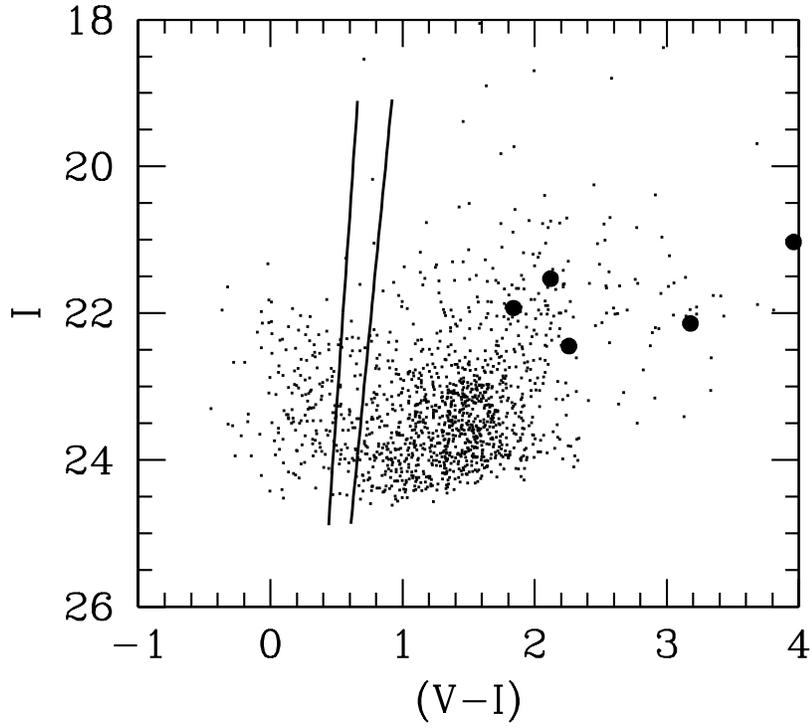,width=16cm}}
\figcaption[cmd_cef.eps]{CMD of DDO187 (same as shown in
Fig. \protect\ref{cmd}) showing (bold dots) the Cepheid stars
candidates identified by Hoessel {\it et al.} (1998). The Cepheids instability
strip is also shown.
\label{cmd_cef}}
\end{figure}

\placefigure{cmd_cef}

In the case of DDO 187, Hoessel {\it et al.} (1998) identified only
two stars as very likely long-period Cepheids and three more, with
lower quality photometry, as possible Cepheids. They used a control
CMD for these stars but perhaps it was not deep enough for the
purpose. Figure \ref{cmd_cef} shows our CMD with the Cepheid
candidates of Hoessel {\it et al.} (1998) marked with filled
dots. From their location in the diagram, these stars are likely AGB
stars and could well be some kind of long-period variables (for
further discussion on the presence of long-period variables in DDO 187
see Hoessel et al. 1998). 

Considering that DDO 187 is the third case in which a large distance
disagreement has been found, it seems clear that Cepheid distance
estimates based on a few stars, as usually happens for dwarf galaxies,
must be accepted with caution. Possibly, only galaxies with a large
young star population, like spirals or dwarfs in a phase immediately
following a strong star formation burst, are adequate for a confident
use of Cepheid stars as distance indicators, since only in these
galaxies a significant population of Cepheids is expected. The
analysis should ever include checking colors in a reliable CMD and a
good phase coverage in order to avoid aliased periods (see also Saha
{\it et al.} 1996 on the distance of the postburst dIrr IC 10).
 
In the following we will use $d_{\rm MW}=2.5\pm 0.2$ Mpc for the
discussion of the properties of DDO 187.

\section{Ionized gas}

\begin{figure}
\centerline{\psfig{figure=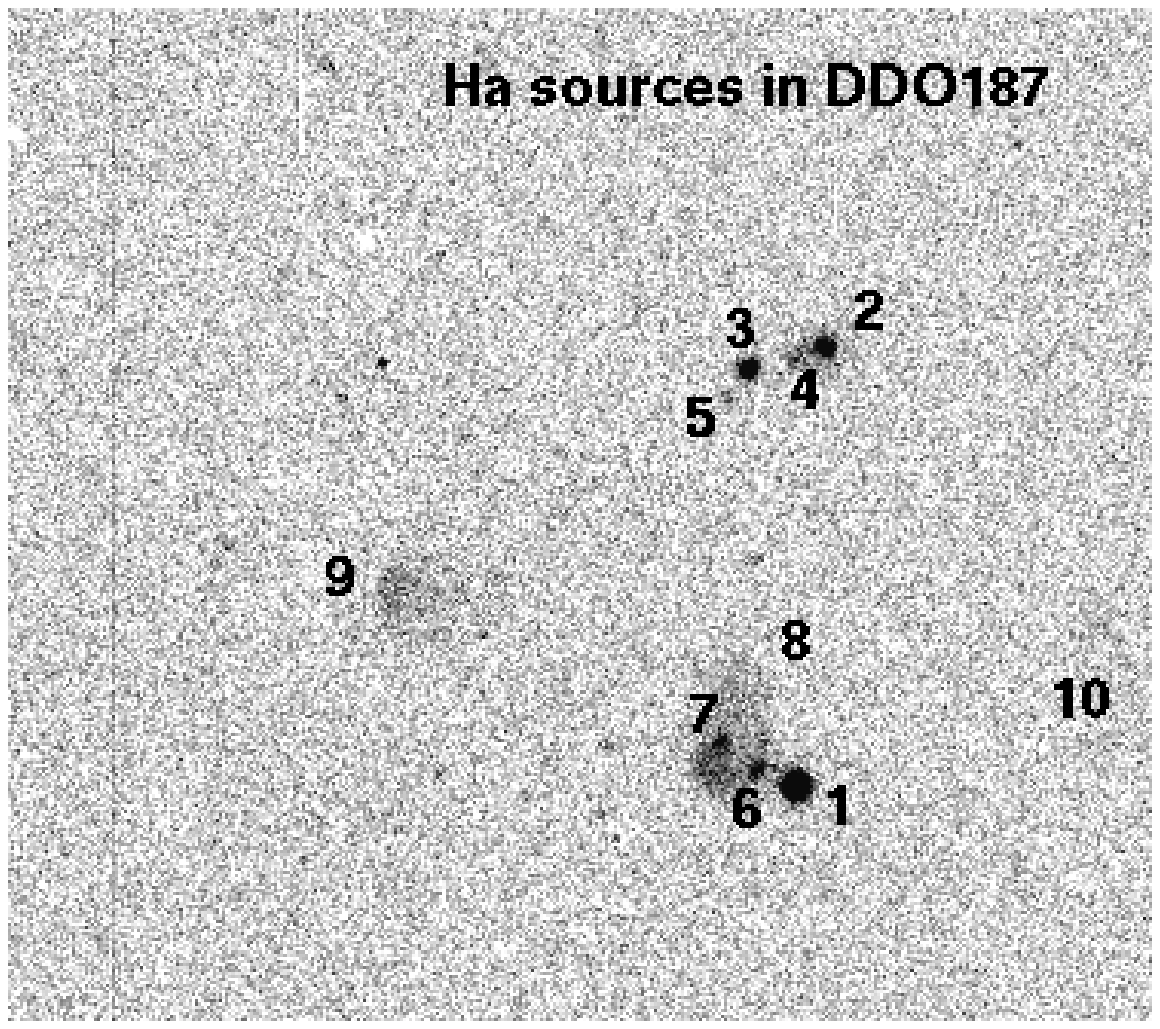,width=16cm}}
\figcaption[ha_labels.ps]{$H_\alpha$ image of DDO 187, after subtraction
of the continuum. The total field is $3.75\times3.75(')^2$ and the
integration time 1200 s. North is up, east is left.
\label{ha_labels}}
\end{figure}

\placefigure{ha_labels}
\placetable{fluxes}

Figure \ref{ha_labels} shows the $H_\alpha$ image of DDO 187, after
subtraction of the continuum. We have identified 10 H~{\sc ii} regions of
different morphology and compactness. A summary of the results is given in
Table \ref{fluxes}. Column 1 gives the identification number as shown in
Fig. \ref{ha_labels}, followed by a bracketed c or d indicating whether the
H~{\sc ii} region is compact or diffuse, respectively. Column 2 lists an
estimate of the size. For regions 1 to 7, the sizes are the full width at
half of maximum, while for the diffuse regions 8 to 10, the sizes are the
square root of the total area used to calculate the flux. Column 3 gives the
flux received from each region in erg s$^{-1}$\,cm$^{-2}$. Column 4 gives the
corresponding $H_\alpha$ luminosities in erg\,s$^{-1}$ calculated for a
distance of 2.5 Mpc. Column 5 lists the number of ionizing photons per second
required to produce those luminosities, calculated following Kennicutt
(1988). Column 6 gives the masses required for single MS stars to produce
such amounts of ionizing photons (they have not been calculated for the
diffuse regions). To calculate these masses, the relation given in Fig. 6 has
been used. It has been plotted from calculations made using the models of
Panaggia (1973) and the Padua stellar evolutionary library (Bertelli {\it et
al.}  1994). It must be kept in mind that, at the distance of DDO 187, $1''$
corresponds to 12 pc, so the assumption that the H~{\sc ii} regions are
ionized by single stars may well be an oversimplification, even for the
compact ones.

\begin{figure}
\centerline{\psfig{figure=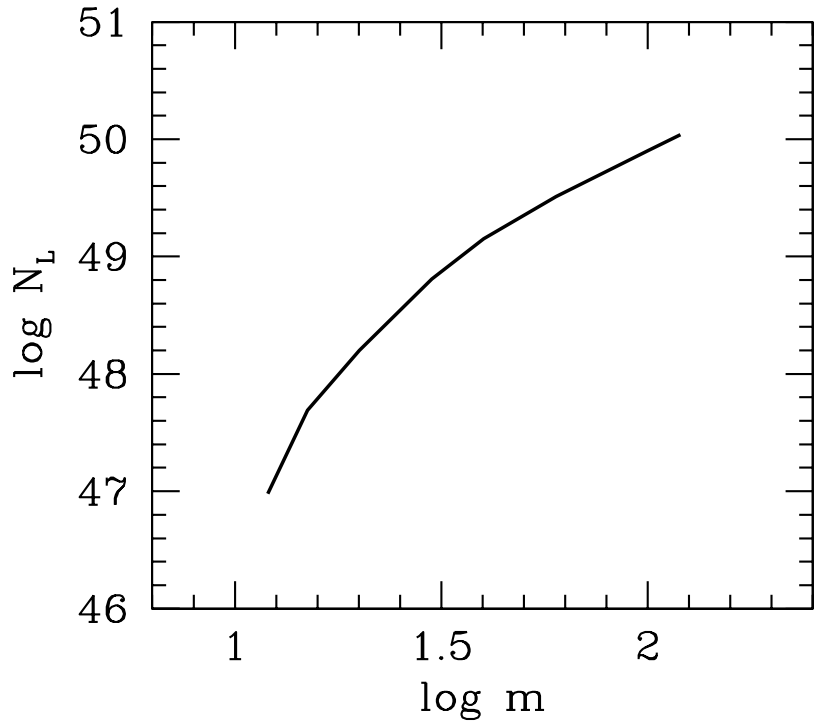,width=16cm}}
\figcaption[lyman.eps]{Relation between mass and Lyman photons flux for
MS stars of metallicity $Z=0.0004$. 
\label{lyman}}
\end{figure}

\placefigure{lyman}

The total $H_\alpha$ flux, luminosity and number of ionizing photons are
given in the last line of Table \ref{fluxes} including also the very diffuse
emission, not comprised in the regions 1 to 10 listed in the table. $N_L$ can
be used to estimate the current SFR. It is given by
\begin{equation}
N_L=\int_{m_k}^{m_s}\int_{0}^{\tau(m)}\phi(m)\,\psi_n(t)\,n_L(m)\,dm\,dt,
\end{equation}
\noindent where time increases from the present
to the past, the present time being 0,
$\phi(m)$ is the IMF, $\psi_n(t)$ is the SFR in units of number of stars
per year and $\tau(m)$ is the MS life-time of a star of mass $m$ and $n_L(m)$
is the number of lyman photons emitted per second by a MS star of mass $m$. While
we are interested in stars more massive than some 10 M$_\odot$ (see Table
2) we need to
consider only the interval of time 0 to $\tau(10)$. If, for simplicity,
we assume that the SFR is constant over this interval, then
\begin{equation}
N_L=\psi_n(0)\int_{m_k}^{m_s}\phi(m)\,n_L(m)\,\tau(m)\,dm,
\end{equation}
\noindent where $\psi_n(0)$ is the constant current value of the SFR in
yr$^{-1}$. Using data for stars of metallicity $Z=0.0004$ from the stellar
evolutionary library of Padua (see Bertelli {\it et al.} 1994) the MS
life-time (in years) can be approximated analytically by
\begin{equation}
\tau(m)=2.24\times10^{8}\,m^{-1}.
\end{equation}
\noindent Moreover, for masses in the interval 12M$_{\sun}\leq m\leq
40$M$_{\sun}$, $n_L(m)$ can be approximated analytically by
\begin{equation}
n_L(m)=10^{48}\,(1.352-0.288\,m+0.015\,m^2).
\end{equation}
Introducing the IMF of Kroupa, Tout, \& Gilmore (1993), equation (4)
becomes
\begin{equation}
N_L=S\times \psi_n(0),
\end{equation}
\noindent with $S=0.91\times 10^{52}$, $1.29\times 10^{52}$, or $1.92\times
10^{52}$ for $m_s=25$M$_{\sun}$, 30M$_{\sun}$ and 40M$_{\sun}$,
respectively. The result is almost insensitive to the value of $m_k$, which
has been fixed to $m_k=12$M$_\odot$. Using the total value of $N_L$ given in
Table \ref{fluxes}, the current SFR can be adopted to be $\psi_n(0)=(3.44\pm
1.22)\times 10^{-3}$yr$^{-1}$. It is better to give the SFR in units of
mass. From integration of the Kroupa {\it et al.} (1993) IMF from 0.1
M$_\odot$ to $\infty$ the average mass of the stars results 0.51 M$_\odot$
and so the SFR becomes $\psi(0)=(1.75\pm 0.62)\times
10^{-3}$M$_{\sun}$yr$^{-1}$.  This can be assumed to be the SFR in DDO 187
averaged for the last $\sim$ 20 Myr.  Different IMFs would produce different
results for the SFR. For comparison, using a Salpeter IMF
($\phi(m)=Am^{-2.35}$) with lower mass limit $m_i=0.1$ M$_\odot$, would
produce a SFR 10\% smaller than the former value.

\section{Morphology, star formation history and structure of DDO 187:
testing the halo/disk hypothesis}

Evident galactocentric gradients in the young stellar population have
recently been found in several dIrrs: WLM (Minniti \& Zijlstra 1996);
Antlia (Aparicio \etal 1997c); Phoenix (Mart\'\i nez-Delgado,
Gallart, \& Aparicio 1999); NGC 3109 (Minniti, Zijlstra, \& Alonso
1999). All these galaxies have undergone recent star formation
episodes in their centers, but lack young stars in outer
regions. However the conclusion cannot be drawn from this fact alone
that they have two differentiated structures, the outer being an old
subsystem tracing out the formation and early evolution of the
galaxy. This result would not be surprising, but sufficiently
accurate age distributions and structural and kinematical information
are needed before confidently accepting it.

Interesting steps towards establishing the primeval nature of the
extended structures are supplied by Minniti, \& Zijlstra (1997) who
provide with clues compatible with an old age of the extended
population in WLM (although the presence of an intermediate-age
component can not be, in our opinion, excluded) and include a
morphological analysis concluding that flattening is smaller for the
halo than for the inner component. A further important step forward
is supplied by the finding that the only globular cluster in WLM is
indeed a very old object, sharing the metallicity of the outer
component field stars (Hodge \etal 1999). However, WLM has also a very
extended gas component (see Minniti, \& Zijlstra 1997) which, at least
and in principle, would play against the formation of the extended
component before the collapse of the galaxy had finished.

Phoenix provides an equally interesting case. Lacking gas associated
to the main, optical body, this galaxy has an outer stellar component
possibly (but not certainly) formed by old stars and rotated
$90\grados$ with respect to the inner component (Mart\'\i nez-Delgado
et al. 1999). This fact, by itself alone, strongly indicates a
possible separation of both substructures.  Together with the presence
of young stars in the inner component and a gas cloud at 6 arcmin away
from them (Young, \& Lo 1997) it makes Phoenix of mayor relevance in
the study of extended components.

Centering our attention on DDO 187, we will show in this section
how our data, together with information about the gas distribution,
could indicate that a stellar population lacking young stars would be
associated with an external, diffuse, perhaps differentiated subsystem
of this galaxy. On the light of these finding, we will discuss the
possible disk-halo structure on the galaxy.

\subsection{The morphology of DDO 187}

From surface brightness photometry, Patterson \& Thuan (1996, 1998) have
determined the Holmberg radius, ellipticity and position angle of DDO
187 to be $r_{26.5}^B=50.4''$, $b/a=0.5$ and PA $=41\grados$. On the other
hand, for the H~{\sc i} component of the galaxy, Lo {\it et al.} (1993)
obtain $a_{\rm HI}=2.0'$, $(b/a)_{\rm HI}=0.8$, PA$_{\rm HI}=130\grados$,
where $a_{\rm HI}$ is the major axis measured to 10\% of the peak
flux. In other words, the gas component has a similar extension 
to that of the
optical light integrated to the Holmberg radius, but, to the first
approximation, the former shows a spherical morphology, in contrast
to the likely flattened structure of the latter. 

To further investigate the structure of the galaxy, we have searched the
morphology of the distribution of stars to its outermost extension. We
have obtained the number density of resolved stars and   fitted it
with ellipses. Position angles and ellipticities for different semi-major
axes have been obtained with the IRAF package. Except for the inner
$45''$, the ellipses show PA in the range 20$\grados$--$50\grados$,
compatible with that of the optical Holmberg radius ellipse, but with
ellipticities $ b/a\sim $ 0.7--0.85. See below for more discussion about
the different components of the galaxy.

\begin{figure}
\centerline{\psfig{figure=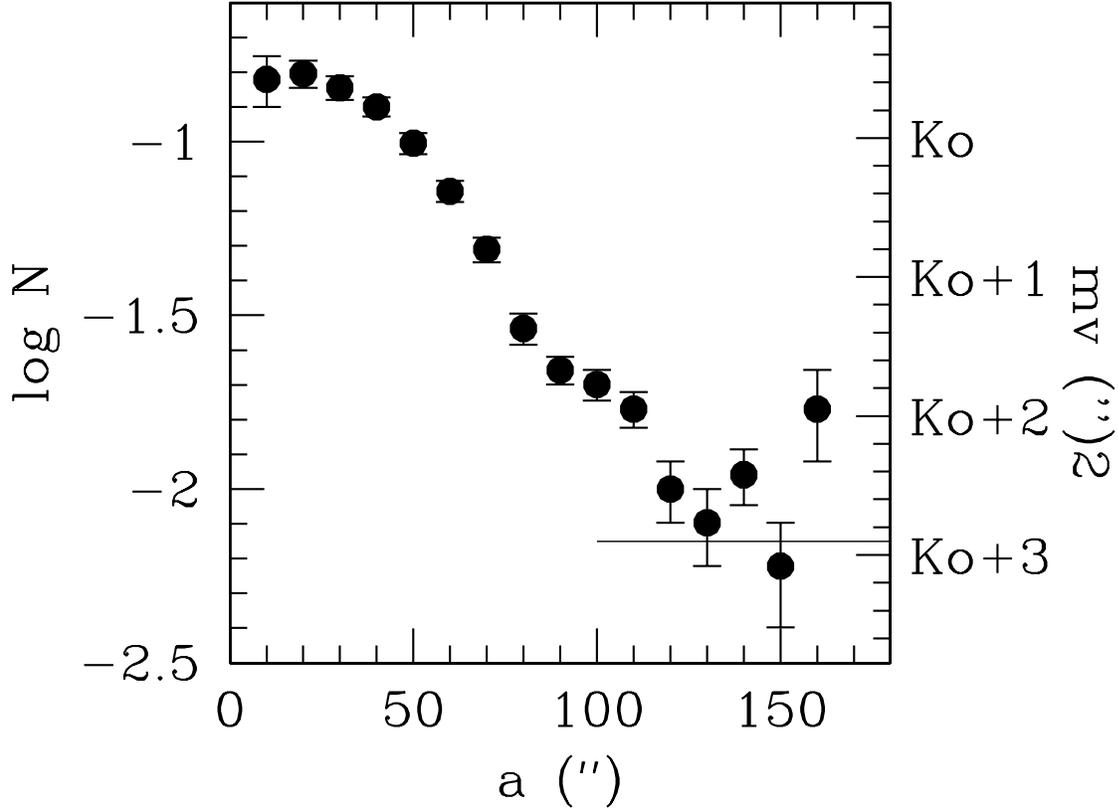,width=16cm}} 
\figcaption[densi.eps]{Radial distribution of
resolved stars in DDO 187. The horizontal axis represents the
semi-major axis distance to the galaxy center. Error bars correspond to
Poisson distribution in the star counts. The line in the lower
right-hand part of the figure represents the background level. The
right-hand vertical axis gives rough photometrical scales for the surface
brightness profiles in $V$ and $I$. The scales can be obtained making
$k_0=24.0$ for $\mu_V$ and $k_0=23.5$ for $\mu_I$. See text for details.
\label{densi}}
\end{figure}

\placefigure{densi}

We then obtained the surface density of stars in the galaxy by
counting stars in ellipses of increasing semi-major axis $r$. All the
stars shown in the CMD of Figure \ref{cmd} have been used. We have
adopted $b/a=0.85$ and PA $=49\grados$ for all the ellipses. These
values correspond to those of the outermost ellipses, at
$r>100''$. The results are shown in Figure \ref{densi}. The background
level estimated from our comparison field is shown by the line in the
lower right part of the figure. After subtracting this background
level, we performed an exponential fit to the region $50''\leq r\leq
120''$. The innermost part of the galaxy has been excluded to avoid
completeness effects. The fit yields a scale factor $\alpha=25.5''$,
which corresponds to 310 pc at the distance of DDO 187. After scaling
to the same distance, this is more than twice the result obtained by
Patterson \& Thuan (1996) for the surface brightness distribution,
indicating that the galaxy is larger than was previously thought. 

In fact, the surface brightness distribution given by Patterson, \&
Thuan (1996) for DDO 187 extends to 50'', i. e. covering only the
inner body of the galaxy. For this reason, roughly estimated surface
brightness scales in $V$ and $I$ are also plotted in the right-hand
vertical axis of Figure \ref{densi}. The scales can be obtained making
$k=24.0$ for the case of $\mu_V$ and $k=23.5$ for the case of $\mu_I$
and have been computed comparing the logarithmic number of stars in a
region of $80\times80 ('')^2$ centered on the galaxy with the
integrated magnitudes of the same region. The surface brightness
profile by Patterson, \& Thuan (1996) is to be preferred for the inner
region of the galaxy, the surface brightness scale of Figure
\ref{densi} being useful for the external region.  Surface magnitude
of the brightest, south-west region of the galaxy, where the youngest
stars are placed are $\mu_V=22.69\pm0.05$ and $22.27\pm 0.05$. Errors
are the zero-point errors of our photometry.

\subsection{The star formation history of DDO 187}

The SFH of a galaxy can be derived in detail from a deep CMD through
comparison with synthetic CMDs (see Gallart {\it et al.} 1999). The CMD
of DDO 187 is not deep enough for such a detailed analysis but it is
still possible to sketch the SFH of the galaxy for old, intermediate and
young ages. 

\begin{figure}
\centerline{\psfig{figure=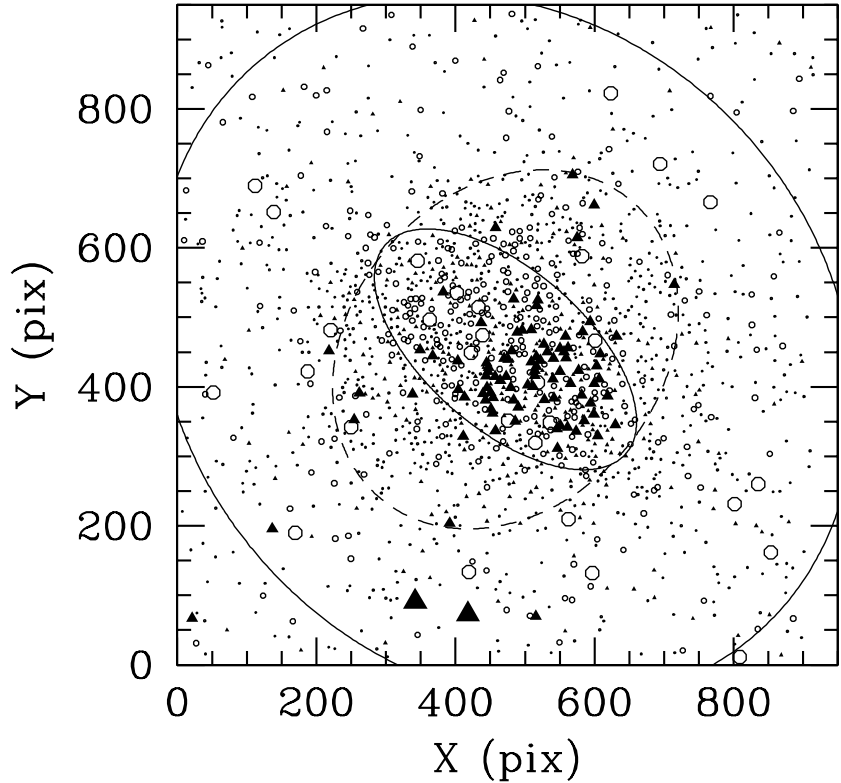,width=16cm}}
\figcaption[mapa.eps]{Resolved stars in DDO 187. Red stars
($(V-I)>0.5$) are represented by open circles. Blue stars ($(V-I)\leq
0.5$) are represented by filled triangles. The three regions in which
the galaxy has been divided for the morphological study are shown by
the over-plotted ellipses. The inner one corresponds to the Holmberg
radius (for $\mu_B\leq 26.5$), the intermediate one corresponds to the
gas extension and the outermost one shows the maximum extension of the
stellar component. Young stars are preferentially located in the
south-west of the inner region. North is up and east is left.
\label{mapa}}
\end{figure}

\placefigure{mapa}

The galaxy has been divided into three parts for the study of the SFH as
shown in Figure \ref{mapa}. The inner part corresponds to the optical body
of the galaxy down to the Holmberg radius, as determined by Patterson
\& Thuan (1996, 1998). The intermediate part matches the gas component
as given by Lo \etal (1993), excluding the inner component. The outer
part includes the rest of the galaxy out to $r=120''$ where, according
to the results given above, the galaxy seems to vanish. The resolved
stars have  also been plotted in Figure \ref{mapa}. While most of the blue
 and many bright red stars are in the inner component, the fainter
red population extends outwards and defines the ellipticity and
orientation of the outer component.

In what follows, we will discuss the derivation of the SFH
for each region. Note that the different stellar populations will evolve
kinematically in different ways and will not necessarily be found in the
same region where they were born. So, properly speaking, what we are going to
obtain is the distribution of stellar ages, with indication of stellar
metallicities in each region. For simplicity, we will continue  to term
 the combined distribution as SFH. We will look for differences in these
distributions as a first indication of differences in the underlying
structure and/or evolutionary history of the galaxy.

A simplified version of the method proposed by Aparicio,
Gallart, \& Bertelli (1997b) and used by Gallart {\it et al.} (1999) for
Leo I has been employed to obtain the SFH of the three regions defined in
DDO 187. In practice, a synthetic CMD with arbitrary, constant SFR of
value $\psi_p$, the IMF of Kroupa {\it et al.} (1993),  25\% of binary
stars with mass ratios, $q$, uniformly distributed in the interval
$0.6<q<1.0$, and a metallicity, $Z$, taking random values from $Z_1=0.0002$
to $Z_2=0.001$ independently of age have been used. The $Z$ interval has
been chosen considering the metallicities for H~{\sc ii} regions given by
Skillman {\it et al.}  (1989) and by van Zee {\it et al.} (1997a), the
value of [Fe/H] given in Sec. 4, and the dispersion of colors in the 
red-tangle.

\begin{figure}
\centerline{\psfig{figure=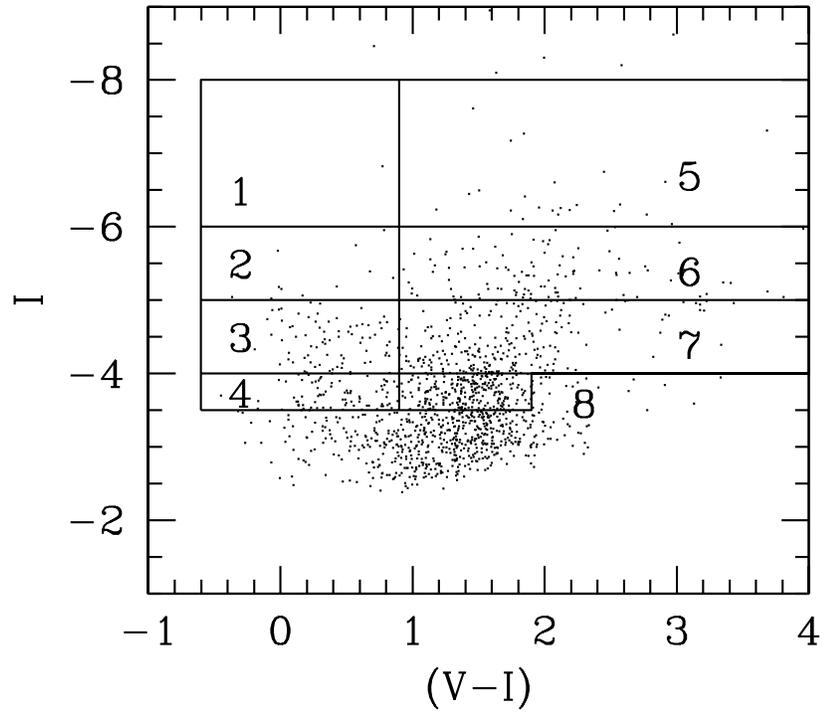,width=16cm}}
\figcaption[cmd_reg.eps]{CMD of DDO187 (same as shown in
Fig. \protect\ref{cmd} after correction of distance modulus
$(m-M)_0=27.0$) showing the eight regions used to parameterize the
distribution of stars of different ages.
\label{cmd_reg}}
\end{figure}

\placefigure{cmd_reg}

The resulting synthetic CMD has been then divided into four age
intervals: 0--0.1 Gyr, 0.1--0.3 Gyr, 0.3--4 Gyr, and 4--15 Gyr. Following the
nomenclature introduced in Aparicio {\it et al.} (1997b), each of these
synthetic diagrams will be called a {\it partial model} CMD and any linear
combination of them will be denoted as {\it global models}. Eight regions
have been defined in the observed
and the partial model CMDs with the criterion that they sample different
age intervals (see Gallart {\it et al.} 1999 and Aparicio 1999). The regions
are shown in Figure \ref{cmd_reg}. We will denote by $N_j^o$ the number of
stars of the observed CMD lying in region $j$ and by $N_{ji}^m$ the number
of stars of partial model (age interval) $i$ populating region
$j$. The number of stars populating a given region in a global model is
then given by
\begin{equation}
N_j^m=k\sum_i\alpha_iN_{ji}^m
\end{equation}
\noindent and the corresponding SFR 
\begin{equation}
\psi(t)=k\sum_i\alpha_i\psi_p\Delta_i(t),
\end{equation}
\noindent where $\alpha_i$ are the linear combination coefficients,
$k$ is a scaling constant, and $\Delta_i(t)=1$ if $t$ is inside the
interval corresponding to partial model $i$ and $\Delta_i(t)=0$
otherwise. The $\psi(t)$ having the best compatibility with the data
can be obtained by a least-squares fitting of $N_j^m$ to $N_j^o$. 

In practice, several solutions can be found showing similar departures,
$\sigma$, from the data. To smooth out the bias that selecting a single
good solution would introduce, several  are selected and averaged
to obtain the best final solution. In the case of DDO 187, models showing
$\sigma$ values smaller than 1.5 times the $\sigma$ of the best solution
have been used.

\begin{figure}
\centerline{\psfig{figure=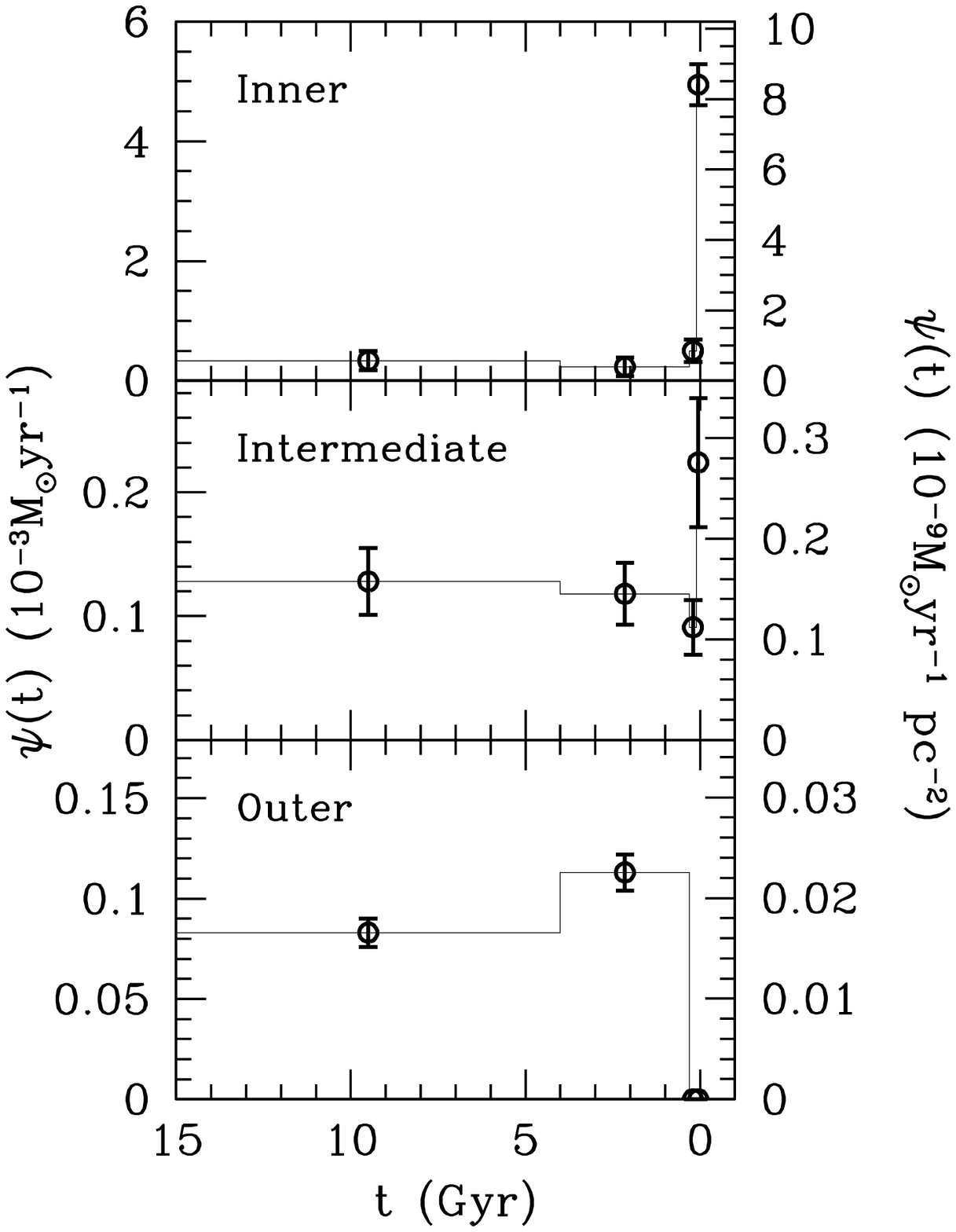,width=16cm}}
\figcaption[psi.eps]{The SFR of DDO 187 as a function of time, obtained
from the analysis of its CMD using synthetic CMDs.
\label{psi}}
\end{figure}

\placefigure{psi}

The resulting $\psi(t)$ for the three regions in which DDO 187 has been
divided are shown in Figure \ref{psi}. The right-hand vertical scales are
normalized to the area covered by each region. 

The most interesting result from the SFH analysis is that the central
region shows a strong increase of $\psi(t)$ for the last few 100 Myr
and that the outer region shows no star formation activity, in good
agreement with what is expected from the absence of gas. This will be
used below for the discussion of the structure of DDO 187. 

The current SFR obtained in Sec. 5 from the H$_\alpha$ luminosity
refers to the last $\sim$ 20 Myr. It is a factor of eight smaller than
the value derived here for the last 100 Myr in the central region of
the galaxy, but it is similar to the SFR averaged over the entire life-time
of the galaxy. This picture favors a bursting star formation mode,
with the last strong burst having taken place between 20 and 100 Myr
ago.  If this burst had lasted 10 Myr, its intensity would have
been of $2\times 10^{-2}$ M$_\odot$ yr$^{-1}$, which is not far from
the values given by Fanelli, O'Connel, \& Thuan (1988) for blue
compact dwarfs.

It is also interesting to note that the picture of high present-day
or very recent star formation activity much higher than the average
for the entire life-time of the galaxy is frequently found in galaxies
classified as dIrr (NGC 6822: Gallart \etal 1996b,c; Pegasus:
Aparicio, Gallart, \& Bertelli 1997a, Gallagher \etal 1999; Antlia:
Aparicio \etal 1997c). On the other hand, galaxies like Phoenix
(Mart\'\i nez-Delgado \etal 1999) or LGS 3 (Aparicio, Gallart {\it et al.}
 1997b), frequently considered as dSph-dIrr transition objects,
show similar $\psi(t)$ histories until a few 100 Myr ago, failing only
in currently producing stars at a high rate.  This would also point to
a bursting mode of star formation, the dSph-dIrr galaxies  being in
a quiescent period, and would also imply that an important bias could
exist in the classification of dwarfs as bona-fide dIrrs towards
objects experiencing strong star formation bursts.

\placetable{sfrave}

The SFR averaged for different periods of time is given in Table
\ref{sfrave} for the three regions defined in DDO 187.  The first four
lines give the SFR averaged for the whole life-time of the galaxy (assumed
to be 15 Gyr) and for the intervals 15--1 Gyr, 1--0 Gyr, and 0.1--0 Gyr
ago. Line 5 gives the current SFR derived from the $H_\alpha$
flux. Lines 6 to 10 give the same information as lines 1 to 5 after
normalization to the area of each region.

\subsection{A disk-halo structure in DDO 187?}

Some of the results presented here are compatible with (but not
conclusive of) a two-components, disk/halo-like structure in DDO
187. Let us assume the two-component possibility as a working hypothesis
and check to what extent it could be accepted. There are three facts
supporting it:

\begin{enumerate}

\item The high ellipticity of the inner star forming region could be
due to it being a flattened, disk-like component, different from the almost
spherical, low-density, outer region.

\item The gas component is spatially much more compact than the outer stellar
one. This would be qualitatively compatible with a collapse that
left behind the formed stars earlier.

\item The external component lacks young stars, as it would be
expected in an early formed structure which was later on deprived of
gas.

\end{enumerate}

These  three results combine both age and morphological
properties, thereby giving more self-consistency to the two-component
hypothesis. But all them are necessary rather than sufficient
conditions. In particular, the sufficient conclusive conditions should be
of the three following kinds:

\begin{enumerate}

\item Kinematical, and not only morphological, decoupling should exist
between the inner and the outer components. Radial velocities of the
stars would be necessary to determine if this is the case, but they
are beyond reach of current instruments.

\item The gas should be associated with the inner, flat component,
rather than showing a spherical distribution.  It could  be, however,
that the sphericity of the gas originates in the heating produced by
supernova explosions in the inner region.

\item The stellar population of the outer component should be old
(older than $\sim$ 10 Gyr), the fact that young stars lack there not
being enough. With the time resolution for intermediate and old ages
of the analysis presented here it is not possible to accept or reject
the presence of an intermediate-age population with a high confidence
level.

\end{enumerate}

Summarizing, the two-component hypothesis does not seem unrealistic.
But nothing can be definitely stated until more detailed data, ideally
including kinematics, are available. By now, this is the case for
other dwarfs for which differentiated components have been sought
(Antlia, NGC 3109 and with stronger clues, WLM and Phoenix; see
references above). For these, as for DDO 187, more data are
needed. 

\section{Global properties of DDO 187}

\placetable{global}

The SFR $\psi(t)$ and the new estimate of the distance obtained here have be
used, together with data from other authors, to calculate the integrated
properties of DDO 187 listed in Table \ref{global}. The bracketed figures
refer to the bibliographic sources for each data. First the metallicity, $Z$,
the distance to the Milky Way, $d_{\rm MW}$, and the distance to the
barycenter of the Local Group, $d_{\rm LG}$, are given. Integrated absolute
magnitudes and luminosities, referred to the Holmberg radius at
$\mu_B=26.5$arcsec$^{-2}$ (which match the inner component of the former
analysis) follow. $M_\star$ is the mass in stars and stellar remnants for the
whole galaxy, calculated from integration of $\psi(t)$ and assuming that 0.8
of this integral remains locked in stellar objects. $M_{\rm gas}$ is obtained
by multiplying the H~{\sc i} mass by 4/3 to take into account the mass in
He. $M_{\rm vt}$ is the virial mass obtained from the velocity dispersion of
the gas. $\mu$ is the gas fraction, relative to the total mass intervening in
the chemical evolution. $dmf$ is the dark matter fraction, calculated as
indicated in the table. Finally, the gas and total mass to luminosity
fractions are given.

\section{Conclusions}

The hypothesis that dwarf star forming galaxies could have extended,
primeval structures is tested for the dIrr DDO 187. Together with the
SFH and the structure of DDO 187, this paper provides a new estimate
of the distance and discusses the distribution of H~{\sc ii} regions.

From the $I$ magnitude of the tip of the red giant branch ($I_{\rm
TRGB}$), the distance of DDO 187 to the Milky Way is estimated to be
$2.5\pm 0.2$ Mpc. The distance to several neighbor galaxies and
groups have been computed, showing that it is an isolated, field
galaxy: DDO 187 is at 2.6 Mpc from M 94; at 2.7 Mpc from the
barycenter of the Local Group, and at 3.2 Mpc from M 81; the closest
dwarf galaxies to DDO 187 are GR 8, at 0.85 Mpc and DDO 190, at 1.05 Mpc.

The distance of DDO 187 to the Milky Way (2.5 Mpc) found here is quite
different from the value of 7 Mpc, estimated from Cepheid light-curves
by Hoessel {\rm et al.} (1998). This is the third case (Pegasus, and
Leo A are the other two) in which such a large disagreement is
obtained between the two methods and implies that Cepheid-based
distances of dwarf galaxies must be accepted cautiously. We suggest
that the scarcity of Cepheids in dwarf galaxies, derived from the
small amount of young stars, would not make worthwhile the effort of
determining their distance with Cepheids.

DDO 187 has several H~{\sc ii} regions. From  analysis of their fluxes, it
is found that the most massive star alive in the galaxy has a mass of some 30
M$_\odot$ and that the current SFR is smaller by a factor of 3 than
it was  20--100 Myr ago.

The SFH of DDO 187 has been obtained using synthetic CMDs. DDO 187
shows an overall decreasing SFR, but with  high star formation
activity over the last few hundred Myr in its central region. The SFR
for different regions of the galaxy are summarized in Table
\ref{sfrave}. In particular, the data analyzed here are compatible
with a maximum in the SFR  between 20 and 100 Myr ago that could
have had an intensity comparable to that of some blue compact
galaxies. Interestingly, this picture of an overall decreasing SFR with a
strong enhancement in the recent past is common to other dIrr
galaxies. It is suggested that perhaps the dIrr morphology is only
produced by the recent past SFR activity and that no large intrinsic
differences exist between typical dIrrs and other dwarf galaxies
showing similar intermediate- to old-age SFHs but neither H~{\sc ii} regions
nor other traces of very recent star formation activity.

Several results suggest that DDO 187 has a two-component
halo/disk-like structure: (i) differentiated morphologies for the
inner and outer stellar components; (ii) a gas component less extended
than the outer stellar component, and (iii) an outer component lacking
young stars. The working hypothesis that a real halo/disk structure
could be present is discussed. The conclusion is reached that the
two-components hypothesis is not unrealistic, but that nothing can be
clearly stated until more detailed data, ideally including
kinematics, are available.

Finally the distance and the SFH derived in this paper have been used
together with data from the literature to calculate several global
parameters of DDO 187 which are given in Table \ref{global}.

\acknowledgments

This paper has been largely improved after careful reading of the
manuscript by Carme Gallart and David Mart\'\i nez-Delgado and after
the endless, fruitful discussions maintained with them. We are also
grateful to Richard Patterson and to the referee of the paper, Albert
Zijlstra, for their useful comments and suggestions.

This work has been financially supported by the IAC (grants P3/94 and
P90/92), by the DGES of the Kingdom of Spain (grant PB97-1438-C02-01)
and by INTAS-RFBR grant 95-IN-RU-1930. Data from NED have been used.

\begin{deluxetable}{ccccc}
\tablenum{1}
\tablewidth{400pt}
\tablecaption{Journal of observations
\label{journal}}
\tablehead{
\colhead{Date} & \colhead{Object} & \colhead{Time (UT)} & \colhead{Filter} &
\colhead{Exp. time (s)}}
\startdata
97.07.26 & DDO 187 & 21:50 & $I$ & 600 \nl
97.07.26 & DDO 187 & 22:01 & $I$ & 600 \nl
97.07.26 & DDO 187 & 22:12 & $V$ & 900 \nl
97.07.26 & DDO 187 & 22:35 & $V$ & 900 \nl
97.07.27 & DDO 187 & 21:46 & $V$ & 900 \nl
97.07.27 & DDO 187 & 22:03 & $V$ & 900 \nl
97.07.27 & DDO 187 & 22:18 & $I$ & 900 \nl
97.07.27 & DDO 187 & 22:35 & $I$ & 900 \nl
97.07.29 & DDO 187 & 21:04 & $H_\alpha$ & 900 \nl
97.07.29 & DDO 187 & 21:19 & $H_\alpha$ & 300 \nl
97.07.29 & DDO 187 & 21:25 & $H_\alpha$-cont & 400 \nl
97.07.29 & Field & 22:20 & $V$ & 600 \nl
97.07.29 & Field & 22:32 & $V$ & 600 \nl
97.07.29 & Field & 22:43 & $I$ & 500 \nl
97.07.29 & Field & 22:51 & $I$ & 500 \nl
\enddata

\end{deluxetable}
\newpage

\begin{deluxetable}{lc}
\tablenum{2}
\tablewidth{250pt}
\tablecaption{Distances of DDO 187 to the closest galaxies and groups
\label{neig}}
\tablehead{
\colhead{Galaxy} & \colhead{$d$ (Mpc)} }
\startdata
Milky Way & 2.5 \nl
Andromeda & 2.9 \nl
Local Group (barycenter) & 2.7 \nl
M 81 & 3.2 \nl
M 94 (CnVI group) & 2.6 \nl
M 101 & 5.2 \nl
GR 8 (dIrr) & 0.85 \nl
DDO 190 (dIrr) & 1.05 \nl
\enddata

\end{deluxetable}
\newpage

\begin{deluxetable}{lcccccccc}
\tablenum{3}
\tablewidth{500pt}
\tablecaption{Cepheid and CMD-based distances of dIrr galaxies
\label{dist}}
\tablehead{
\colhead{Galaxy} & \colhead{\em} & \colhead{$d_{\rm Cep}$} &
\colhead{$n_{\rm Cep}$} & \colhead{Ref.} & \colhead{\em} 
& \colhead{$d_{\rm CMD}$} & \colhead{method} & \colhead{Ref.} }
\startdata 

DDO 187 & & $7\pm 1.2$ & 2 & (6) & & $2.5\pm 0.2$ & TRGB & (1) \nl 
Leo A (DDO 69) & & $2.2\pm 0.2$ & 4 & (5) & & $0.69\pm 0.06$ & RC & (7) \nl
 & & & & & & $0.80\pm 0.08$ & TRGB & (7) \nl 
Pegasus (DDO 216) & & $1.75\pm 0.15$ & 5 & (4) & & $0.95\pm 0.05$ & TRGB & (2) \nl 
 & & & & & & $0.76\pm 0.10$ & TRGB & (3) \nl

\enddata

\tablerefs{(1) This paper; (2) Aparicio (1994); (3) Gallagher {\it et
al.} (1998); (4) Hoessel {\it et al.} (1990); (5) Hoessel {\it et al.}
(1994); (6) Hoessel {\it et al.} (1998); (7) Tolstoy {\it et al.} (1998)}

\end{deluxetable}

\newpage
\begin{deluxetable}{cccccc}
\tablenum{4}
\tablewidth{400pt}
\tablecaption{H~{\sc ii} regions in DDO 187
\label{fluxes}}
\tablehead{
\colhead{\#} & \colhead{$D$} & \colhead{$F_{H\alpha}$} &
\colhead{$L_{H\alpha}$} & \colhead{$N_L$} & \colhead{$M_\star$}
\nl
\colhead{~} & \colhead{$('')$} & \colhead{$10^{-15}$erg\,s$^{-1}$\,cm$^{-2}$} &
\colhead{$10^{36}$erg\,s$^{-1}$} & \colhead{$10^{48}$s$^{-1}$} & \colhead{M$_\odot$}}
\startdata
1 (c) & 4.7 & 13.79 & 10.32 & 7.53 & 31.5 \nl
2 (c) & 3.8 & 4.10 & 3.07 & 2.23 & 22.0 \nl
3 (c) & 4.1 & 3.16 & 2.37 & 1.73 & 20.5 \nl
4 (c) & 6.3 & 0.58 & 0.43 & 0.31 & 14.0 \nl
5 (c) & 5.6 & 0.41 & 0.30 & 0.22 & 13.5 \nl
6 (c) & 5.1 & 0.74 & 0.55 & 0.40 & 14.5 \nl
7 (c) & 6.7 & 0.82 & 0.61 & 0.44 & 14.8 \nl
8 (d) & 39.2 & 21.45 & 16.06 & 11.72 & \nodata \nl
9 (d) & 27.5 & 6.69 & 5.00 & 3.65 & \nodata \nl
10 (d) & 18.5 & 0.58 & 0.43 & 0.31 & \nodata \nl
Total & \nodata & 78.2 & 58.5 & 42.7 & \nodata \nl
\enddata

\end{deluxetable}

\begin{deluxetable}{lccc}
\tablenum{5}
\tablewidth{400pt}
\tablecaption{Summary of the Star Formation History of DDO 187
\label{sfrave}}
\tablehead{ \colhead{\hfill Regions:} & \colhead{Inner} &
\colhead{Intermediate} & \colhead{Outer}} 
\startdata 

$\bar\psi_{15-0}$\hfill ($10^{-4}$M$_\odot$yr$^{-1}$) \hspace {12mm} &
3.4 & 1.3 & 0.9 \nl 
$\bar\psi_{15-1}$ \hfill ($10^{-4}$M$_\odot$yr$^{-1}$)
\hspace {12mm} & 3.1 & 1.3 & 0.9 \nl 
$\bar\psi_{1-0}$ \hfill ($10^{-4}$M$_\odot$yr$^{-1}$) \hspace {12mm} &
7.6 & 1.2 & \nodata \nl 
$\psi_{0.1-0}$ \hfill ($10^{-4}$M$_\odot$yr$^{-1}$) \hspace {12mm} &
50 & 2.2 & 0 \nl 
$\psi(0)_{\rm HII}$ \hfill ($10^{-4}$M$_\odot$yr$^{-1}$) \hspace {12mm} &
17.5 & 0 & 0 \nl 
$\bar\psi_{15-0}/A$ \hfill ($10^{-9}$M$_\odot$yr$^{-1}$pc$^{-2}$)
\hspace {12mm} & 0.59 & 0.16 & 0.018 \nl 
$\bar\psi_{15-1}/A$ \hfill ($10^{-9}$M$_\odot$yr$^{-1}$pc$^{-2}$)
\hspace {12mm} & 0.53 & 0.16 & 0.018 \nl 
$\bar\psi_{1-0}/A$ \hfill ($10^{-9}$M$_\odot$yr$^{-1}$pc$^{-2}$) \hspace
{12mm} & 1.29 & 0.15 & \nodata \nl 
$\psi_{0.1-0}/A$ \hfill ($10^{-9}$M$_\odot$yr$^{-1}$pc$^{-2}$) \hspace
{12mm} & 8.41 & 0.28 & 0 \nl 
$\psi(0)_{\rm HII}$ \hfill ($10^{-4}$M$_\odot$yr$^{-1}$) \hspace {12mm} &
2.93 & 0 & 0 \nl 

\enddata
\end{deluxetable}

\begin{deluxetable}{lc}
\tablenum{6}
\tablewidth{350pt}
\tablecaption{Global properties of DDO 187
\label{global}}
\tablehead{\colhead{} & \colhead{}}
\startdata 

$\alpha_{2000}$, $\delta_{2000}$ & $\rm 14^h 15^m 56^s$, $23\grados 03' 13''$ \nl
$l, b$ & $25\grados.6, +70\grados.5$ \nl
$Z$ & 0.0008 (6,7) \nl 
$d$ \hfill (Mpc) & 2.5 (1) \nl 
$d_{\rm LG}$ \hfill (Mpc) & 2.7 (1) \nl 
$M_{V,0}$ & --13.07 (5) \nl 
$M_{I,0}$ & --13.28 (4) \nl
$L_B$ \hfill ($10^6$L$_\odot$) & 21 \nl 
$L_V$ \hfill ($10^6$L$_\odot$) & 15 \nl 
$L_I$ \hfill ($10^6$L$_\odot$) & 8.5 \nl 
$M_\star$ \hfill ($10^6$M$_\odot$) & 8.4 (1) \nl 
$M_{\rm gas}$ \hfill ($10^6$M$_\odot$) & 26 (2,3,8) \nl 
$M_{\rm vt}$ \hfill ($10^6$M$_\odot$) & 102 (2,3) \nl
$\mu=M_{gas}/(M_\star+M_{gas})$ ~~~ & 0.76 \nl
$dmf=1-(M_\star+M_{\rm gas})/M_{\rm vt}$ & 0.66 \nl 
$M_{\rm gas}/L_V$ \hfill (M$_\odot$/L$_\odot$) & 1.7 \nl 
$M_{\rm vt}/L_V$ \hfill (M$_\odot$/L$_\odot$) & 6.8 \nl 

\enddata

\tablerefs{(1) This paper; (2) Hutchmeier, \& Richter (1986); (3) Lo
{\it et al.} (1993); (4) Patterson, \& Thuan (1996); (5) Prugniel, \&
H\'eraudeau (1998); (6) Skillman {\it et al.} (1989); (7) van Zee {\it
et al.} (1997a); (8) van Zee {\it et al.} (1997b)}

\end{deluxetable}

\newpage

\begin{references}

\reference{} Aparicio, A. 1994, \apjl, 437, L27
\reference{} Aparicio, A. 1999, in The Stellar Content of Local Group
Galaxies. I.A.U. symp. 192, edited by P. Whitelock
\reference{} Aparicio, A., Dalcanton, J. J., Gallart, C., \& Mart\'\i
nez-Delgado, D. 1997c, \aj, 114, 1447
\reference{} Aparicio, A., \& Gallart, C. 1994, in The Local Group:
Comparative and Global Properties, ESO Conference and Workshop
Proceedings No. 51, edited by A. Layden, R. C. Smith, and J. Storm (ESO
Garching), p. 115
\reference{} Aparicio, A., \& Gallart, C. 1995, \aj, 110, 2105
\reference{} Aparicio, A., Gallart, C., \& Bertelli, G. 1997a, \aj, 114, 669
\reference{} Aparicio, A., Gallart, C., \& Bertelli, G. 1997b, \aj, 114, 680
\reference{} Aparicio, A., Gallart, C., Chiosi, C., \& Bertelli,
G. 1996, \apjl, 469, L97
\reference{} Aparicio, A., Moles, A., \& Garc\'\i a-Pelayo, J. M. 1988,
\aaps, 74, 367
\reference{} Aparicio, A., Tikhonov, N., \& Karachentsev, I. 1999, in
preparation
\reference{} Bertelli, G., Bressan, A., Chiosi, C., Fagotto, F., \&
Nasi, E. 1994, \aaps, 106, 275
\reference{} Burstein, D., \& Heiles, C. 1982, \aj, 87, 65
\reference{} Dohm-Palmer, R. C., Skillman, E. D., Gallagher, J.,
Tolstoy, E., Mateo, M., Dufour, R. J., Saha, A., Hoessel, J., \&
Chiosi, C. 1998, \aj, 116, 1227
\reference{} Fanelli, M. N., O'Connel, R. W., \& Thuan, T. X. 1988,
\apj, 334, 665
\reference{} Freedman, W. L., \& Madore, B. F. 1990, \apj, 365, 186
\reference{} Freedman, W. L., Hughes, S. M., Madore B. F., Mould,
J. R., Lee, M. G., Stetson, P, Kennicutt, R. C., Turner, A.,
Ferrarese, L., Ford, H., Graham, J. A., Hill, R., Hoessel, J. G.,
Huchra, J., Illingworth, G. D. 1994, \apj, 427, 628
\reference{} Gallagher, J. S., Tolstoy, E., Dohm-Palmer, R. C.,
Skillman, E. D., Cole, A. A., Hoessel, J. G., Saha, A., \& Mateo,
M. 1998, \aj, 115, 1869
\reference{} Gallart, C., Freedman, W.L., Aparicio, A., Bertelli, G., \&
Chiosi, C. 1999, \aj, submitted
\reference{} Gallart, C., Aparicio, A., Bertelli, G., \& Chiosi,
C. 1996b, \aj, 112, 1950
\reference{} Gallart, C., Aparicio, A., Bertelli, G., \& Chiosi,
C. 1996c, \aj, 112, 2596
\reference{} Gallart, C., Aparicio, A., \& V\'\i lchez, J. M. 1996a, \aj,
112, 1928
\reference{} Hodge, P. W., Dolphin, A. E., Smith, T. R., \& Mateo, M.,
1999, \apj, in press
\reference{} Hoessel, J. G., Abbott, M. J., Saha, A., Mossman, A. E.,
\& Danielson, G. E. 1990, \aj, 100, 1151
\reference{} Hoessel, J. G., Saha, A., \& Danielson, G. E. 1998, \aj, 116,
1679
\reference{} Hoessel, J. G., Saha, A., Krist, J., \& Danielson,
G. E. 1994, \aj, 108, 645
\reference{} Kennicutt, R. C. 1988, \apj, 334, 144
\reference{} Kroupa, P., Tout, C. A., \& Gilmore, G. 1993, \mnras, 262, 545
\reference{} Landolt, A. U. 1992, \aj, 104, 340
\reference{} Lee, M. G., Freedman, W. L., \& Madore, B. F. 1993, \apj,
417, 553
\reference{} Lo, K. Y., Sargent, W. L. W., \& Young, K. 1993, \aj, 106,
507
\reference{} Mart\'\i nez-Delgado, D., Gallart, C., \& Aparicio,
A. 1999, \aj, in press
\reference{} Minniti, D., Zijlstra, A. A. 1996, \apj, 467, L13
\reference{} Minniti, D., Zijlstra, A. A. 1997, \aj, 114, 147
\reference{} Minniti, D., Zijlstra, A. A., \& Alonso, M. V. 1999, \aj, 117, 881
\reference{} Mulder, P. S., \& van Driel, W. 1993, \aa, 272, 63
\reference{} Oke, J. B. 1990, \aj, 99, 1621
\reference{} Panagia, N. 1973, \aj, 78, 929
\reference{} Patterson, R. J., \& Thuan, T. X. 1996, \apjs, 107, 103
\reference{} Patterson, R. J., \& Thuan, T. X. 1998, \apjs, 117, 633
\reference{} Peebles, P. J. E. 1989, \apj, 344, L53
\reference{} Saha, A., Hoessel, J. G., Krist, J., \& Danielson,
G. E. 1996, \aj, 111, 197
\reference{} Skillman, E. D., Kennicutt, R. C., \& Hodge, P. H. 1989,
\apj, 347, 875
\reference{} Stetson, P. B. 1994, \pasp, 106, 250 
\reference{} Stetson, P. B., Saha, A., Ferrarese, L., {\it et al.}
1998, \apj,508, 491
\reference{} Tolstoy, E., Gallagher, J. S. Cole, A. A., Hoessel, J. G.,
Saha, A., Dhom-Plamer, R. C., Skillman, E. D., Mateo, M., \&
Hurley-Keller, D. 1998, \aj, 116, 1244
\reference{} van den Bergh, S. 1959,  Pub. David Dunlap Observatory,
vol. II, n. 5
\reference{} van Zee, L., Haynes, M. P., \& Salzer, J. J. 1997a, \aj,
114, 2479
\reference{} van Zee, L., Maddalena, R. J., Haynes, M. P., Hogg, D. E.,
\& Roberts, M. S. 1997b, \aj,
113, 1638
\reference{} White, S. D. M., \& Rees, M. J. 1978, \mnras, 183, 341
\reference{} Young, L. M., \& Lo, K. Y. 1997, \apj, 490, 710

\end{references}
\end{document}